\documentclass[letterpaper, 10 pt, conference]{ieeeconf}  

\IEEEoverridecommandlockouts                              

\usepackage{graphics} 
\usepackage{epsfig} 
\usepackage{times} 
\usepackage{amsmath} 
\usepackage{amssymb}  
\let\proof\relax 
\let\endproof\relax
\usepackage{amsthm}
\usepackage[T1]{fontenc}
\usepackage{soul}
\usepackage{url}
\usepackage[hidelinks]{hyperref}
\usepackage[utf8]{inputenc}
\usepackage[small]{caption}
\usepackage{booktabs}
\usepackage{algorithm}
\usepackage[noend]{algpseudocode}
\usepackage{algorithmicx}
\usepackage[switch]{lineno}
\usepackage{subcaption}
\usepackage{cleveref}

\usepackage[dvipsnames]{xcolor}

\algnewcommand\algorithmicinput{\textbf{Input:}}
\algnewcommand\INPUT{\item[\algorithmicinput]}
\algnewcommand\algorithmicoutput{\textbf{Output:}}
\algnewcommand\OUTPUT{\item[\algorithmicoutput]}

\title{\LARGE \bf
Accelerating Focal Search in Multi-Agent Path Finding\\with Tighter Lower Bounds
}

\author{Yimin Tang$^{1}$, Zhenghong Yu$^{2}$, Jiaoyang Li$^3$, Sven Koenig$^{4}$
\thanks{$^{1}$Thomas Lord Department of Computer Science, University of Southern California 
        {\tt\small yimintan@usc.edu}}%
\thanks{$^{2}$University of Wisconsin–Madison 
        {\tt\small zyu379@wisc.edu}}%
\thanks{$^{3}$Carnegie Mellon University 
        {\tt\small jiaoyanl@andrew.cmu.edu}}%
\thanks{$^{4}$University of California, Irvine 
        {\tt\small sven.koenig@uci.edu}}%
}%

\begin{document}

\maketitle
\thispagestyle{empty}
\pagestyle{empty}

\begin{abstract}

Multi-Agent Path Finding (MAPF) involves finding collision-free paths for multiple agents while minimizing a cost function—an NP-hard problem. Bounded suboptimal methods like Enhanced Conflict-Based Search (ECBS) and Explicit Estimation CBS (EECBS) balance solution quality with computational efficiency using focal search mechanisms. While effective, traditional focal search faces a limitation: the lower bound (LB) value determining which nodes enter the FOCAL list often increases slowly in early search stages, resulting in a constrained search space that delays finding valid solutions. In this paper, we propose a novel bounded suboptimal algorithm, double-ECBS (DECBS), to address this issue by first determining the maximum LB value and then employing a best-first search guided by this LB to find a collision-free path. Experimental results demonstrate that DECBS outperforms ECBS in most test cases and is compatible with existing optimization techniques. DECBS can reduce nearly 30\% high-level Constraint Tree (CT) nodes and 50\% low-level focal search nodes. When agent density is high, DECBS achieves a 23.5\% average runtime improvement over ECBS with identical suboptimality bounds and optimizations.

\end{abstract}

\section{INTRODUCTION}

Automated warehouses represent a multi-billion-dollar industry dominated by companies. These facilities employ hundreds of robots to transport goods between locations~\cite{wurman2008coordinating}. A key aspect of their operation is designing collision-free plans for these robots, a challenge that can be formulated as a Multi-Agent Path Finding (MAPF) problem~\cite{stern2019multi}. MAPF focuses on finding collision-free paths for multiple agents moving from their start locations to designated goal locations in a known environment while optimizing a specified cost function. Many optimal algorithms have been proposed to solve this problem, including $M^*$~\cite{wagner2011m}, Conflict-Based Search (CBS)~\cite{sharon2015conflict}, ICBS~\cite{boyarski2015icbs}, and CBSH2-RTC~\cite{li2019improved}.

Solving the MAPF problem optimally is known to be NP-hard~\cite{yu2013structure}, creating significant challenges for optimal solvers regarding scalability and efficiency. In contrast, suboptimal MAPF solvers—including Prioritized Planning (PP)~\cite{erdmann1987multiple,silver2005cooperative}, PBS~\cite{ma2019searching}, and their variants~\cite{chan2023greedy,li2022mapf}—offer improved computational performance but lack the theoretical guarantees on solution quality. There also exists asymptotically optimal method such as LaCAM~\cite{okumura2023lacam}. Bounded-suboptimal approaches like Enhanced CBS (ECBS)~\cite{barer2014suboptimal} and EECBS~\cite{li2021eecbs} provide a trade-off between efficiency and solution quality, guaranteeing collision-free solutions' costs are within a user-defined suboptimality factor of the optimal cost.

The key idea behind bounded-suboptimal MAPF solvers lies in relaxing the best-first search requirement while incorporating additional heuristic functions to select the next node for expansion effectively. ECBS~\cite{barer2014suboptimal} achieves bounded suboptimality by replacing CBS's best-first search strategy at both the high and low levels of CBS with focal search~\cite{pearl1982studies,cohen2018anytime}. Focal search operates through two queues: an OPEN list containing all unexpanded nodes sorted by cost, a FOCAL list containing a subset of these nodes. The algorithm selects expansion candidates from the FOCAL list using an auxiliary heuristic to accelerate finding bounded-suboptimal solutions. In our implementation, we use collision minimization as the FOCAL heuristic function at both search levels, explicitly employing the negative count of collisions in paths or complete solutions. During focal search, since the optimal cost is unknown until an optimal solution is found, focal search uses the lowest cost in the OPEN list as a lower bound (LB) value to estimate the optimal cost and define FOCAL list eligibility. However, when the LB value is small, the FOCAL list contains fewer nodes, causing focal search to behave similarly to standard best-first search. This constraint also leads to unbalanced search tree expansion~\cite{li2021eecbs}.

In this paper, we introduce double-ECBS (DECBS), a novel bounded-suboptimal variant of ECBS. DECBS modifies the original focal search used in low-level path planning by incorporating an optimized search manner, what we call \textit{double search}. This approach is inspired by ITA-ECBS~\cite{tang2024ita}, a bounded-suboptimal version of ITA-CBS~\cite{tang2023solving}. However, ITA-ECBS primarily focuses on integrating ECBS and ITA-CBS, and it has not been developed as a universal optimization applicable to other methods. The double search operates in two phases: first, a shortest path search determines the optimal cost, the upper bound for the LB value; second, DECBS employs a best-first search guided by the FOCAL list auxiliary heuristic function using this LB value to find a path with fewer collisions. Importantly, paths returned during the low-level search are not necessarily collision-free—they need only to satisfy the constraints specified in the corresponding high-level Constraint Tree (CT) node. 

This maximum LB value offers three benefits: (1) by accurately reflecting the optimal cost, it enables more nodes to enter FOCAL list. This increases the likelihood of selecting a path with fewer collisions, thereby reducing both high-level node expansion and low-level node expansion; (2) The node of focal search is more expensive than \(A^*\) because it requires computing an additional heuristic. In double search, we first perform an \(A^*\) search to obtain a larger LB. This larger LB, in turn, reduces the number of nodes generated by focal search. Although \(A^*\) search introduces additional nodes, these nodes are cheaper to compute than focal search nodes. (3) Candidate nodes with costs exceeding the suboptimality can be pruned, saving memory during search operations. In summary, although DECBS performs two path searches for each low-level search, double search is faster than executing a single focal search. Our empirical evaluation confirms these benefits, showing DECBS reduces nearly 30\% high-level CT nodes and 50\% low-level focal search nodes, and achieves a 23.5\% average runtime improvement over ECBS with identical optimizations when agent density is high.


\section{Problem Definition}\label{sec:problem_definition}

The Multi-Agent Path Finding (MAPF) problem is defined as follows: Given a set of agents $I=\{1,2,\cdots,N\}$ and an undirected graph $G = (V,E)$, where vertices $v \in V$ represent possible agent locations and edges $e \in E$ represent unit-cost movements between two locations. Self-loop edges are allowed which represent ``wait-in-place'' actions. Each agent $i\in I$ has a unique start location $s_i \in V$ and a unique goal location $g_i \in V$, such that $s_i \neq s_j$ and $g_i \neq g_j$ for all $i \neq j$. The task is to find collision-free paths for all agents $i \in I$ from their start locations $s_i$ to their goal locations $g_i$.

All agent actions--including waiting in place and moving to an adjacent vertex--take one time unit.
Let $v^i_t \in V$ be the location of agent $i$ at timestep $t$, and $\pi_i=[v_0^{i}, v_1^{i}, ..., v_{T^{i}}^{i}]$ denote a path of agent $i$ from its start location $v_0^{i}$ to its target $v_{T^{i}}^{i}$. 
We assume that agents rest at their targets after completion, i.e., $v_t^i = v_{T^i}^i, \forall t > T^i$.
The cost of agent $i$'s path is defined as $T^i$, with the minimum cost path referred to as the shortest path.

We consider two types of agent-agent collisions: \emph{vertex collisions}, where two agents occupy the same vertex simultaneously. 
and \emph{edge collision}, where two agents move in opposite directions along the same edge.
We denote both collision types as $(i, j, t)$, a vertex collision between agents $i$ and $j$ at timestep $t$ or an edge collision between agents $i$ and $j$ at timestep $t$ to $t+1$.
The collision-free requirement implicitly requires that all agent goals must be distinct.

Formally, the objective of the MAPF problem is to find a set of paths $\{\pi_i | i\in I\}$ for all agents that satisfies the following conditions:
\begin{enumerate}
\item Each agent $i$ begins at its start location (i.e., $v_0^{i} = s_i$) and reaches its target location $g_j$ (i.e., $v_{t}^{i} = g_j, \forall t \ge T^{i}$);
\item Every pair of adjacent vertices on path $\pi_i$ is connected by an edge (i.e., $(v_{t}^{i}, v_{t+1}^{i}) \in E, \forall 0 \leq t \le T^i$);
\item The complete solution is collision-free while minimizing the total \emph{flowtime} $\sum_{i=1}^{N}T^{i}$.
\end{enumerate}

\section{Related work}\label{sec:related_work}

\subsection{Focal Search}

Focal search is a bounded-suboptimal search algorithm that balances efficiency and solution quality~\cite{pearl1982studies,cohen2018anytime}. Given a user-defined suboptimality factor \(w \geq 1\), the algorithm guarantees a solution \(\pi\) with cost \(c^{val}\) satisfying \(c^{val} \leq w \cdot c^{opt}\), with \(c^{opt}\) representing the optimal solution cost. To achieve this, focal search maintains two queues: OPEN and FOCAL.\footnote{Since OPEN and FOCAL are used for sorting and FOCAL is a subset of OPEN, candidate node pointers are typically stored in both. If a node appears in both queues, only one instance is maintained, with two pointers referencing it.}

The OPEN list stores all candidate nodes awaiting expansion, sorted by the function \(f(n) = g(n) + h(n)\), where \(g(n)\) is the cost-so-far and \(h(n)\) is an admissible heuristic estimate of cost-to-go, similar to \(A^*\) search. The FOCAL queue contains only those nodes \(n\) from OPEN satisfying \(f(n) \leq w \cdot f_{front}\), where \(f_{front}\) is the minimum \(f\) value in OPEN. Nodes in FOCAL are sorted according to an additional heuristic function \(d(n)\) that guides the search toward promising regions of the search space. Focal search always expands the front node in the FOCAL list. When a solution with cost \(c^{val}\) is found, the algorithm sets the lower bound (LB) \(c^g=f_{front}\). We can observe that \(f_{front} \leq c^{opt}\) and \(f_{front} \leq c^{val} \leq w f_{front} \leq w c^{opt}\). Finally, focal search returns two outputs: the LB \(c^g\) and the solution with cost \(c^{val}\). If no solution exists, both \(c^g\) and \(c^{val}\) are set to \(\infty\).

\subsection{Multi-Agent Path Finding (MAPF)}

Multi-Agent Path Finding (MAPF) has a long history~\cite{silver2005cooperative}. Many algorithms have been developed to solve it or its variants. Decoupled algorithms~\cite{silver2005cooperative,luna2011push,wang2008fast} plan paths for each agent independently and combine these paths into a single solution. Coupled algorithms~\cite{standley2010finding,standley2011complete} plan simultaneously for all agents. There also exist dynamically-coupled algorithms~\cite{sharon2015conflict,wagner2015subdimensional,okumura2023lacam} initially generate individual paths and re-plan for multiple agents together when resolving collisions. Among these approaches, Conflict-Based Search (CBS)~\cite{sharon2015conflict} is a basic centralized optimal MAPF algorithm. Some bounded-suboptimal algorithms, such as ECBS~\cite{barer2014suboptimal} and EECBS~\cite{li2021eecbs}, are based on it.

\paragraph{CBS}
Conflict-Based Search (CBS) is an optimal two-level search algorithm. 
The low-level computes shortest paths for individual agents from their start locations to goals, while the high-level searches a binary Constraint Tree. 
Each CT node \(H = (c, \Omega, \pi)\) includes a constraint set \(\Omega\), a solution \(\pi\) containing shortest paths satisfying \(\Omega\) for all agents, and the solution cost \(c\). When a solution \( \pi \) or a path does not include any agent actions or positions that are restricted by a \(\Omega\), we say this solution or path satisfies the \(\Omega\). Importantly, a solution \(\pi\) satisfying \(\Omega\) may still contain collisions; we consider \(\pi\) a valid solution only when it is collision-free.

When expanding a node \(H\), CBS selects a collision in \(H. \pi\) and formulates two constraints, each prohibiting one agent from either occupying the colliding location or performing the conflicting action at the colliding timestep. We have two types of constraints: vertex constraint $(i, v, t)$ that prohibits agent $i$ from occupying vertex $v$ at timestep $t$ and edge constraint $(i, u, v, t)$ that prohibits agent $i$ from moving from vertex $u$ to vertex $v$ at timestep $t$. CBS then generates two successor nodes identical to \(H\), adding one of the new constraints to each successor node's constraint set. After adding a new constraint, the affected agent's apth must be re-planned to satisfy the updated constraint set. By maintaining a priority queue OPEN ordered by solution cost, CBS repeats this process until expanding a collision-free node, which represents the optimal valid solution. CBS guarantees optimality for flowtime minimization~\cite{sharon2015conflict}.

\paragraph{BCBS and ECBS} 
BCBS~\cite{barer2014suboptimal} and ECBS~\cite{barer2014suboptimal} build upon CBS by incorporating focal search. In ECBS low-level search, focal search returns both an LB value \(c^g_i\) and a valid path \(\pi_i\), with cost \(c_i\) for agent $i$, satisfying: \(c^g_i \leq c_i^{opt} \leq c_i \leq w \cdot c^g_i\), where \(c_i^{opt}\) represents optimal path cost of agent \(i\) under constraints \(\Omega\). At the high-level search, ECBS extends CT node structure to \(H = (c, \Omega, \pi, L, c_L)\), adding an array \(L\), that stores all LB values $c^g_i$ for paths in \(\pi\), and cost \(c_L\) represents the sum of these lower bounds. ECBS's high-level search maintains two priority queues: FOCAL and OPEN. OPEN sorts all CT nodes by ascending \(c_L\), while FOCAL contains nodes \(H\) from OPEN satisfying \(H.c \leq w \cdot H_{front}.c_L\), where \(H_{front}\) is the front CT node in OPEN. FOCAL is ordered by a user-defined heuristic function \(d(H)\). ECBS guarantees that returned solution \(H^{sol}.\pi\) satisfies \(H^{sol}.c \leq w \cdot c^{opt}\), with \(c^{opt}\) being the optimal valid solution cost. BCBS similarly employs focal search at both levels but defines the high-level CT node LB value as the sum of the low-level path costs. BCBS is typically denoted as BCBS\((w_1, w_2)\), where \(w_1\) and \(w_2\) are the suboptimality factors for high- and low-level search. When \(w_1w_2 = w\), BCBS\((w_1, w_2)\) returns a solution that is bounded-suboptimal with factor \(w\). Notably, BCBS\((w, 1)\) is equivalent to using \(A^*\) in low-level and focal search in high-level.

\begin{figure}[tbp]
\centering
\includegraphics[width=0.43\textwidth]{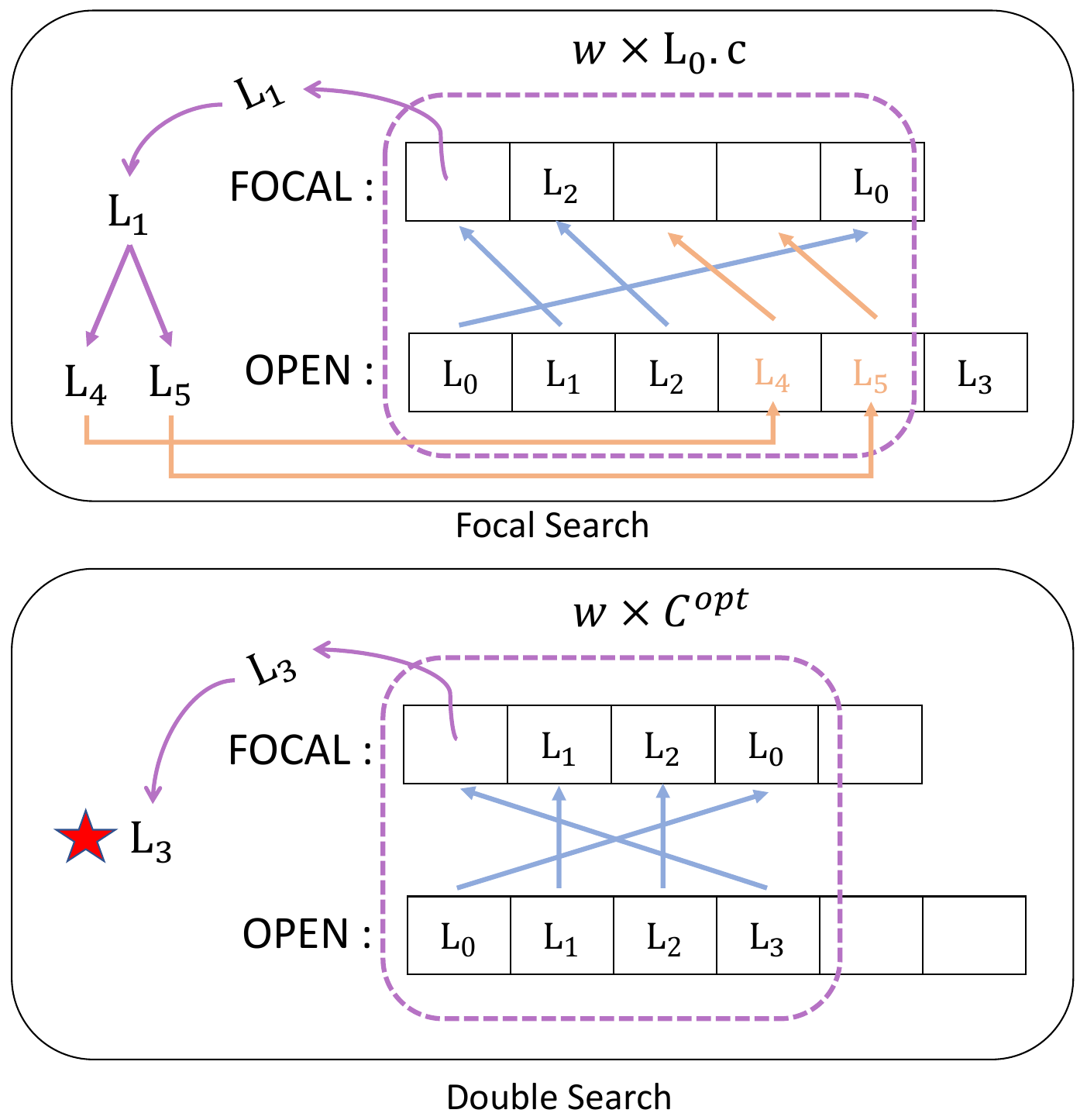} 
\caption{A ideal case to explain how double search is faster than focal search: Let's assume \(L_i\) are different low-level path candidate nodes, and \(L_0.c = L_1.c = L_2.c = 10\), \(L_4.c = L_5.c = 11\), \(L_3.c = 12\) and \(c^{opt} = 11\). We also assume \(w = 1.1\), \(L_3\) is the goal candidate node and the others are not. We have \(d(L_0) < d(L_5) \leq d(L_4) \leq d(L_2) \leq d(L_1) < d(L_3) = 0\) in negative collision numbers. \textbf{Focal Search}: Since the LB value is \(L_0 = 10\) and \(w = 1.1\), all initial nodes \(L_0\), \(L_1\), and \(L_2\) are included in FOCAL except \(L_3\). Then \(L_1\) is selected to expand and we obtain two new nodes \(L_4\) and \(L_5\). When \(L_4\) and \(L_5\) are inserted into OPEN, they also go into FOCAL. As \(L_0\) has the lowest \(d(L_0)\) value, the LB value cannot be improved and \(L_3\) cannot be considered until \(L_0\) is expanded and a new, larger LB value is reached. In practice, there could be a large number of nodes between \(L_0\) and \(L_1\) in FOCAL. \textbf{Double Search}: Here we assume we have already obtained \(c^{opt} = 11\) from the shortest path search. With this information, all nodes are included in FOCAL. Because \(L_3\) has the largest \(d(L_3)\) value, it will be the first node to be searched.} 
\label{fig:algo_overview}
\end{figure}

\section{Method}\label{sec:method}

In this section, we present a detailed description of DECBS. While DECBS maintains focal search for high-level CT node exploration, it diverges from ECBS in low-level path finding. Instead of using focal search to obtain both the lower bound \(c^g\) and solution cost \(c^{val}\), DECBS employs a novel approach we call \textit{double search}, to obtain these values.

To understand the motivation behind double search, let’s first consider ECBS low-level focal search. As previously described, when focal search finds a valid path with cost \(c^{val}\), it establishes the lower bound as \(c^g = f_{front}\). This relationship satisfies:
\[
f_{front} \leq c^{opt} \leq c^{val} \leq w f_{front} \leq w c^{opt}
\]

During focal search, nodes from OPEN with costs in the range \([f_{front}, w f_{front}]\) are selected into the FOCAL. If we want to include more nodes in FOCAL (e.t. potentially find solutions more quickly), we must increase \(f_{front}\) by expanding the front node in OPEN. However, this process is often inefficient because \(f_{front}\) increases slowly--it only changes after exhausting all existing and potential child nodes with the current minimum cost. 

We observe that a more effective range would be \([f_{front}, w· c^{opt}]\), as indicated by the inequality. Since \(c^{opt}\) represents an upper bound for \(f_{front}\), this range maximizes the number of candidates for FOCAL selection while still maintaining the theoretical suboptimality guarantee. The challenge of using \([f_{front}, wc^{opt}]\) is that $c^{opt}$ remains unknown during focal search. Determining $c^{opt}$ requires running an optimal path finding algorithm like \(A^*\). So in our double search approach, we first run shortest path search such as $A^*$ to obtain $c^{opt}$. Once we have $c^{opt}$, we no longer need OPEN from focal search, as we can immediately discard any node with cost exceeding $wc^{opt}$, since they won't be in a  bound-suboptimal solution. This allows us to conduct a best-first search exclusively on the FOCAL, using conflict number as the heuristic function to sort all CT nodes.  

\Cref{alg:DECBS} presents the pseudo-code for DECBS. The key component appears in Lines 23–26: first, we determine the lower bound \(c^g\) through a shortest path search, then use this value to filter out nodes exceeding the suboptimality bound before adding the remaining nodes into the FOCAL. Finally, we expand nodes from FOCAL according to the the FOCAL heuristic function.

\begin{algorithm}[t!]
\small
\caption{\small DECBS}
\label{alg:DECBS}
\textbf{Input}: Graph \(G\), start locations $\{s_i\}$, target locations $\{g_i\}$, suboptimality factor $w$\\
\textbf{Output}: A valid MAPF solution within the suboptimality factor $w$ 

\begin{algorithmic}[1] 
\State \(H_{0}\) = new CTnode()
\State \(H_{0}.\Omega\) = $\emptyset$
\State FOCAL = OPEN = PriorityQueue()
\State Calculate \(d(H_0)\) and insert \(H_{0}\) into OPEN
\While{OPEN not empty}
\State \(H_{front}\) = OPEN.front() 
\State FOCAL = FOCAL $\cup$ \{H $\in$ OPEN $\mid$ $H.c \leq w \cdot H_{front}.c$\}
\State \(H_{cur}\) = FOCAL.front(); FOCAL.pop()
\State Delete \(H_{cur}\) from OPEN
\If {\(H_{cur}.\pi\) has no collision}
    \State \textbf{return} \(H_{cur}.\pi\)
\EndIf
\State ($i, j, t$) = getFirstCollision($H_{cur}.\pi$)
\For{\textbf{each} agent $k$ in ($i,j$)}
    \State $Q$ = a copy of \(H_{cur}\)
    \If{($i, j, t$) is a vertex collision}
        \State \(Q.\Omega\) = \(Q.\Omega\) $\cup$ ($k$, $v^k_{t}$, $t$) // vertex constraint
    \Else 
        \State \(Q.\Omega\) = \(Q.\Omega\) $\cup$ ($k$, $v^k_{t-1}$, $v^k_{t}$, $t$) // edge constraint
    \EndIf
    \(Q.\pi, Q.L\) = lowLevelSearch($G, s_k, g_x, Q, w$)
    \State \(Q.c_L\) = sum(\(Q.L\))
    \If{Q.c $\textless$ $\infty$}
        \State Calculate \(d(Q)\) and insert $Q$ into OPEN
    \EndIf
\EndFor
\EndWhile
\State \textbf{return} No valid solution

\Function{lowLevelSearch}{$G$, $s_k$, $g_x$, $Q, w$}
    \State \(c^g\) = shortestPathSearch($G$, $s_k$, $g_x$, \(Q.\Omega\))
    \State \(c^{val}\) =  bestFirstSearch($G$, $s_k$, $g_x$, \(Q.\Omega\), $w$, \(c^g\))
\State \textbf{return} \(c^g\), \(c^{val}\)
\EndFunction
\end{algorithmic}
\end{algorithm}

People may have two questions, since running  an optimal search followed by a bounded-suboptimal search seems wasteful and inefficient--(1) Why perform two searches when the first already yields a valid path?
(2) Is double search more efficient than single focal search?

 Actually, this apparent redundancy is precisely the insight behind our \textit{double search}. The answer for the first question is straightforward: we deliberately seek a suboptimal low-level path rather than an optimal one because suboptimal paths often introduce fewer collisions in subsequent planning stages, improving overall algorithm efficiency.

For the second question, although performing a double search might seem slower than a single search, three factors make our approach more efficient:
(1) By obtaining \(c^{opt}\), we can include more nodes in the FOCAL list compared to ECBS. A larger FOCAL set increases the likelihood of finding a bounded-suboptimal path with fewer conflicts, thereby reducing the number of nodes expanded during the low-level search. We show an idea example in \Cref{fig:algo_overview} to explain this situation.
(2) Since we already have an upper bound for each agent's path LB value, the CT node LB becomes larger. This larger LB provides the high-level focal search with greater flexibility in selecting FOCAL nodes for expansion, resulting in fewer CT nodes overall.
(3) Focal search is more computationally expensive than \(A^*\) because it requires evaluating an additional heuristic. The larger LB reduces the number of nodes generated by the focal search. Although \(A^*\) search introduces extra nodes, these nodes are cheaper to compute than those produced by focal search. In effect, the additional \(A^*\) search helps offset the cost of extra nodes, leading to an overall reduction in computation time.

\begin{figure}[tbp]
  \centering
  \begin{subfigure}[b]{0.39\textwidth}
    \centering
    \includegraphics[width=\textwidth]{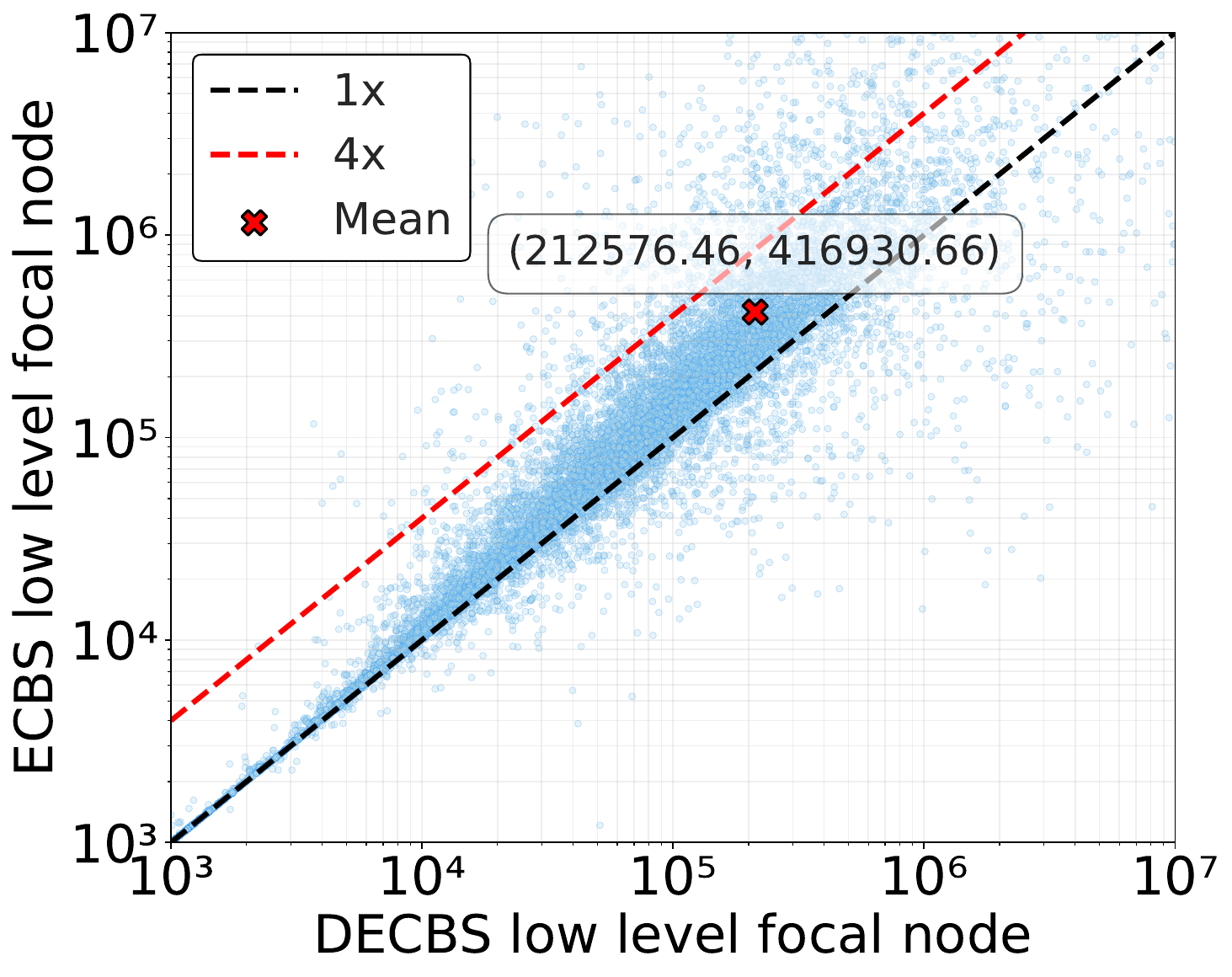}
    \label{fig:low}
  \end{subfigure}
  \quad
  \begin{subfigure}[b]{0.39\textwidth}
    \centering
    \includegraphics[width=\textwidth]{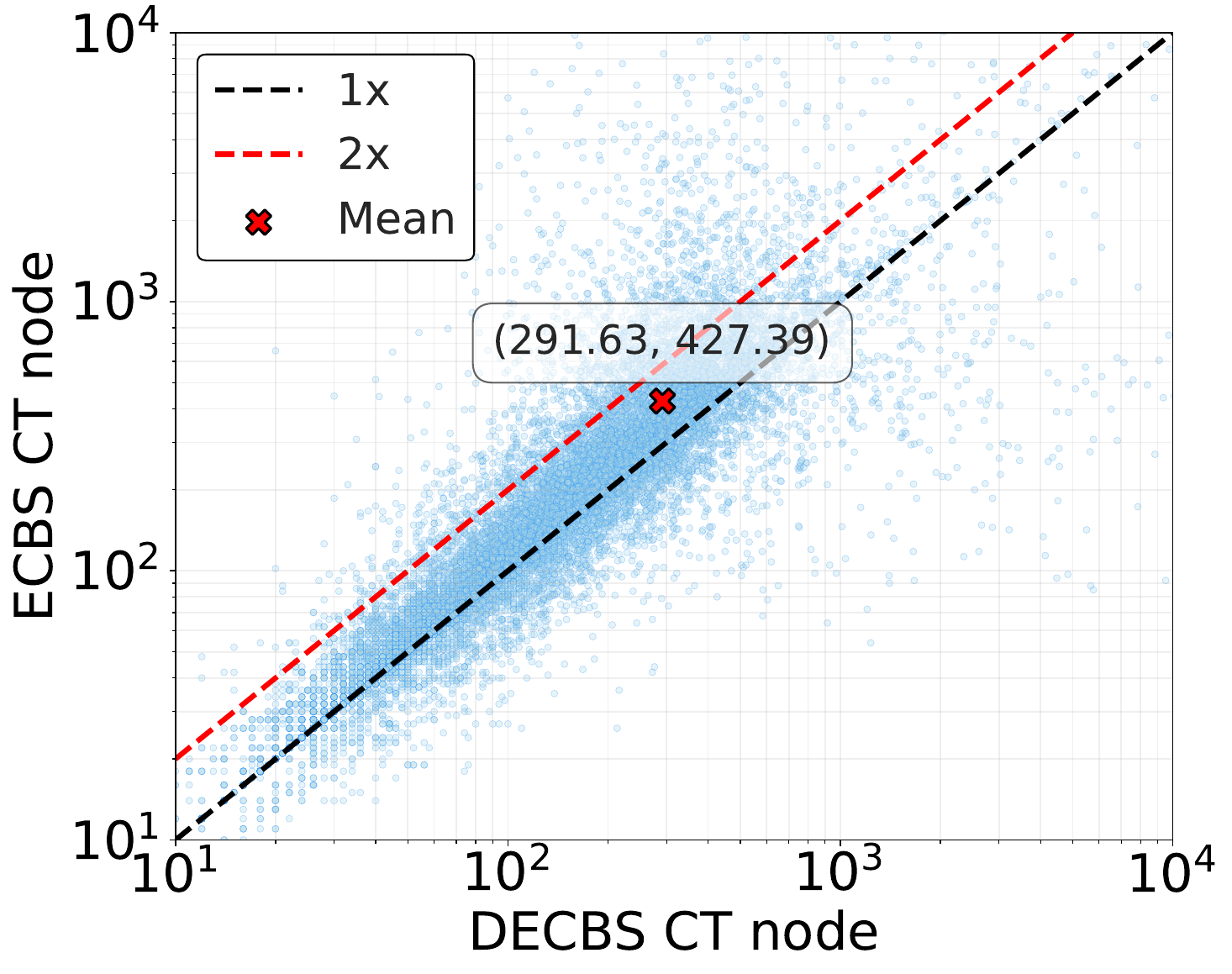}
    \label{fig:high}
  \end{subfigure}
  \caption{Comparison of Search Nodes between DECBS and ECBS: Points above the blue line (1x) indicate that DECBS expands fewer search nodes than ECBS. Only test cases where both algorithms finished within the time limit were counted. The top figure shows the number of low-level search nodes—focal search expanded nodes for ECBS and best-first search nodes for DECBS. The bottom figure displays the number of expanded CT nodes.}
  \label{fig:combined}
\end{figure}

\begin{figure*}[htbp]
\centering
\includegraphics[width=\textwidth]{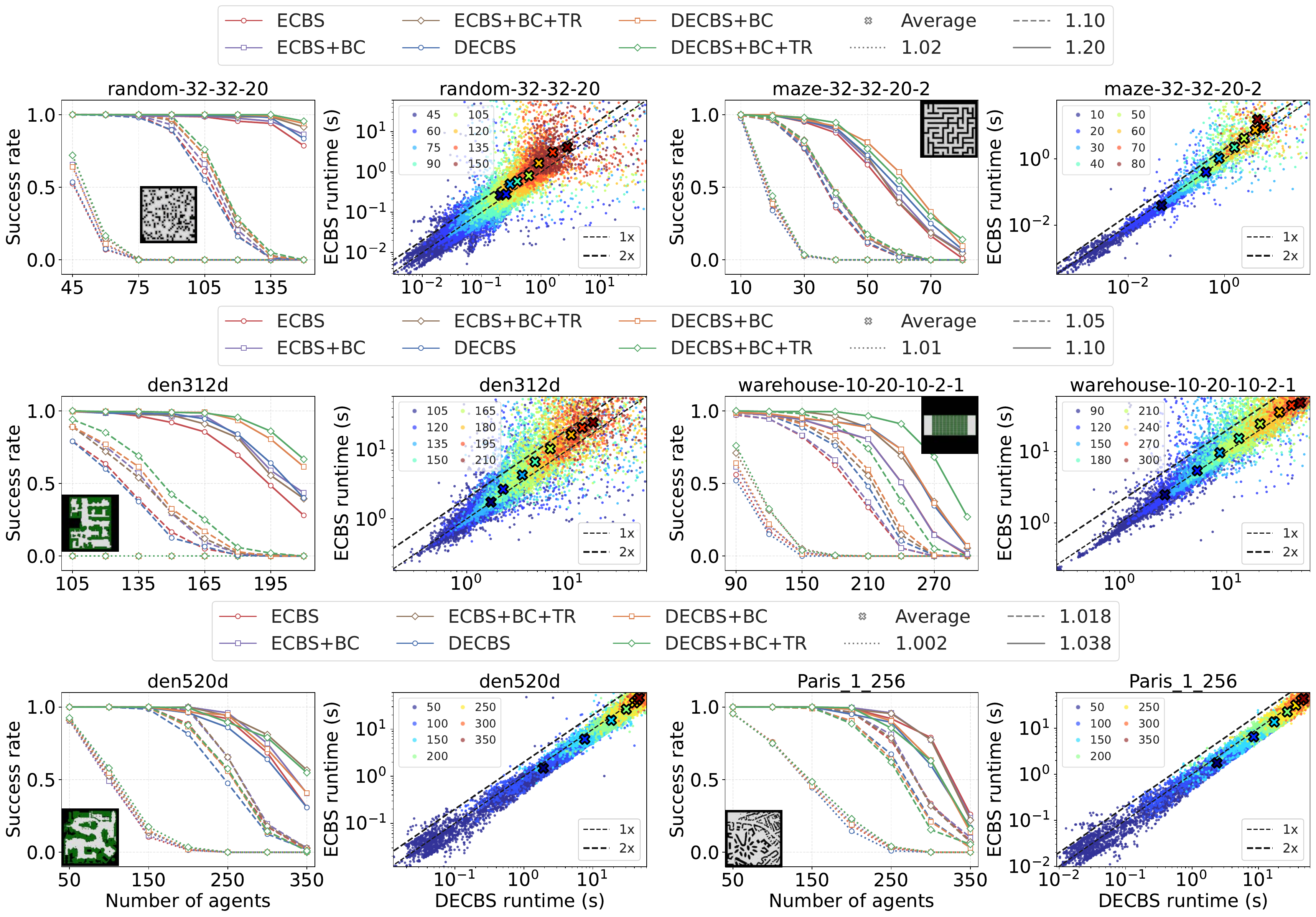} 
\caption{Success rate and runtime: The first and third column subfigures show the success rates of DECBS and ECBS with BC and TR optimizations across different agent numbers. The second and fourth column subfigures compare the runtimes of DECBS and ECBS (here DECBS includes DECBS, DECBS+BC and DECBS+BC+TR, and similarly for ECBS). In the runtime subfigures, color represents the number of agents in the test cases, and the cross mark indicates the average value.}
\label{fig:overview}
\end{figure*}

\section{Experimental Results}\label{sec:experimental_result}

In our experiments, we evaluate DECBS against ECBS, including variants with two optimization techniques, Bypassing Conflicts (BC)~\cite{boyarski2015don} and Target Reasoning (TR)~\cite{li2021pairwise}.
We implement DECBS and ECBS in Rust.\footnote{
Our code is already available at \url{https://github.com/HarukiMoriarty/RUST-CBS}.} To ensure a fair comparison, both DECBS and ECBS share the same code base for core components such as low-level focal search and high-level search node expansion.
All experiments are conducted on an Ubuntu 20.04.1 system with an AMD Ryzen 3990X 64-Core Processor with 2133 MHz 196GB RAM.

\begin{figure*}[tbp]
\centering
\includegraphics[width=\textwidth]{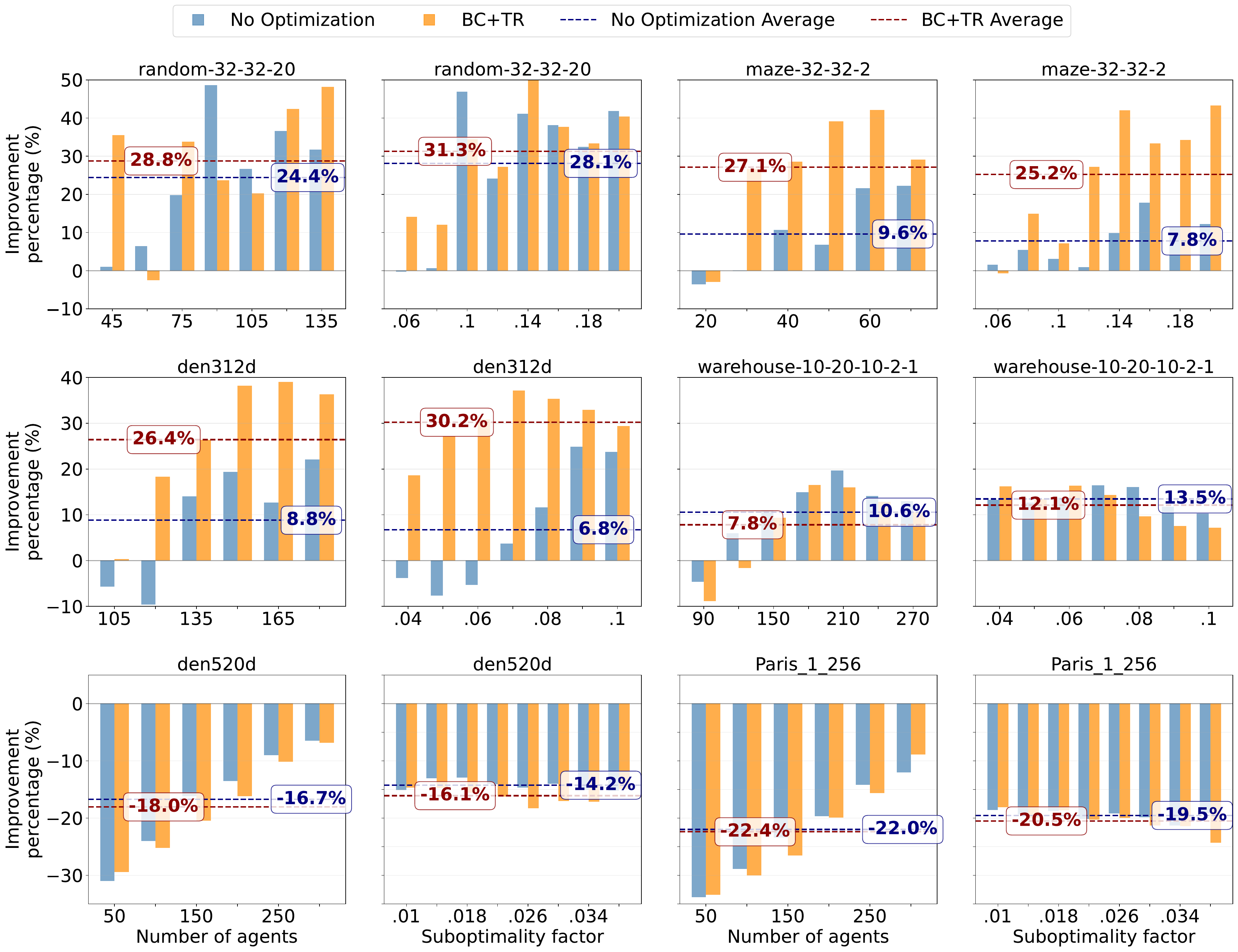} 
\caption{In this figure, we show average DECBS runtime improvements over ECBS on the same test cases and using the same optimization methods. We calculate improvement as follows: \((runtime(ECBS)-runtime(DECBS))/runtime(ECBS)\). Columns 1 and 3 represent improvements for different agent numbers, while columns 2 and 4 represent improvements for different suboptimality factors. We only selected test cases that were solved by both algorithms.}
\label{fig:improvement}
\end{figure*}

\subsection{Test Settings}

We evaluated DECBS and ECBS on 6 maps from the MAPF Benchmark~\cite{stern2019mapf}, as shown in \Cref{fig:overview}: (1) random-32-32-20 (32x32) and maze-32-32-20 (32x32) are grid maps with random obstacles and designed obstacles, (2) den\_312d (65x81) is a grid map from the video game Dragon Age Origins, (3) warehouse-10-20-10-2-1 (161x63) is a grid map inspired by real-world autonomous warehouses, and (4) den\_520d (256x257) and Paris\_1\_256 (256x256) are large benchmark maps. We use map-specific suboptimality factors: random-32-32-20 and empty-32-32-20 (1.02 to 1.20, interval 0.02), den\_312d and warehouse-10-20-10-2-1 (1.01 to 1.10, interval 0.01), den\_520d and Paris\_1\_256 (1.002 to 1.038, interval 0.004). For agent numbers, we have empty-32-32-20 (10 to 80, interval 10),  random-32-32-20 (45 to 150, interval 15), den\_312d (105 to 210, interval 15), warehouse-10-20-10-2-1 (90 to 300, interval 30), den\_520d and Paris\_1\_256 (100 to 450, interval 50). For each combination of map, agent number and suboptimality factor, we generate 200 random test cases, resulting in total 270,000 test cases. An algorithm is considered to have failed for a given test case if it does not find a valid solution within 60 seconds (with runtime recorded as 60 seconds). Success rate is defined as the percentage of test cases solved successfully within the time limit under specific test settings.

\subsection{Performance}

We first show double search reduces the number of low-level nodes in FOCAL and high-level CT nodes. As shown in \Cref{fig:combined}, each data point in the figure represents a test case that both DECBS and ECBS completed within the time limit. On average, DECBS only searches 212k nodes in the low-level FOCAL compared to ECBS's 416k nodes, achieving nearly 50\% node savings. For high-level CT nodes, DECBS expands only 291 nodes compared to 427 for ECBS, resulting in nearly 30\% node savings. Although DECBS performs an additional shortest path search in the low-level search, \Cref{fig:overview} (columns 2 and 4) shows that the total runtime of DECBS for test cases is lower than that of ECBS, indicating that the time saved by reducing FOCAL nodes and CT nodes more than compensates for the extra shortest path search cost, especially in high agent density scenarios.

\Cref{fig:overview} shows detailed success rates over different maps, numbers of agents, suboptimality factors, and various optimization methods. In the column 1 and 3 subfigures of \Cref{fig:overview}, we can observe that DECBS outperforms ECBS in success rate on all small and medium maps. In particular, on the warehouse map, there is a noticeable gap between the green line for DECBS+BC+TR and the other lines. Also, on the den\_312d map with \(w=1.1\), DECBS outperforms ECBS+BC+TR, and all three variants—DECBS, DECBS+BC, and DECBS+BC+TR—consistently outperform the other methods under different conditions. As for the column 2 and 4 subfigures of \Cref{fig:overview}, we plot the runtimes of the two algorithms using different colors to indicate the number of agents. It is clear that as the number of agents increases, the acceleration effect becomes more pronounced. As agent density grows, it becomes harder for focal search to find a low-cost goal state. In contrast, with double search, we obtain the highest LB value, which provides a better chance to jump over unpromising nodes and devote more time to nodes with fewer conflicts.

However, in \Cref{fig:overview}, on large maps such as den\_520d and Paris\_1\_256, we can observe that DECBS is slower than ECBS. This is mainly because, compared to the map sizes, the number of agents is too small. With such low agent density, collisions between agents rarely occur. In this case, focal search behaves more like \(A^*\) search, and double search essentially becomes two \(A^*\) searches. However, we can also observe in the runtime subfigures that as the number of agents increases, the crossover point moves further and further above the 1x line.

We present the average speedup of DECBS over ECBS for different agent numbers and suboptimality factors in \Cref{fig:improvement}. Overall, DECBS is 23.5\% faster than ECBS on average with BC+TR optimization in small and medium maps, and as the number of agents increases, the improvements become more pronounced. Moreover, when both algorithms use the same optimizations (such as BC+TR), the optimization method yields even more performance gains for DECBS. Also, as the suboptimality factor increases, DECBS exhibits more improvements compared to ECBS. However, on large maps, DECBS is slower than ECBS. Still, we observe that as the number of agents increases, the gap between DECBS and ECBS shrinks. These large map results support our earlier statements: in large maps with low agent density, focal search degenerates into \(A^*\) search, which means DECBS essentially performs two \(A^*\) searches. That is why, when the number of agents increases, the speed of DECBS improves and approaches that of ECBS because collisions become more frequent. Regarding suboptimality, the performance degradation remains similar because, in large maps, there are hardly any collisions between agents. In such cases, the shortest path is likely collision-free, making the FOCAL search heuristic ineffective.

\section{Conclusion}\label{sec:conclusion}

In this paper, we have proposed a novel bounded suboptimal algorithm called double-ECBS (DECBS), which replaces traditional focal search with a two-phase approach consisting of a shorted path search followed by a best-first search. Our extensive experiments demonstrate that this double search optimization significantly reduces both low-level FOCAL nodes and high-level CT nodes, particularly in high agent density scenarios. DECBS also shows compatibility with other optimization methods and achieves a higher speedup compared to ECBS with identical optimizations. Looking forward, we find the core principle behind DECBS has potential applications beyond MAPF, we plan to explore the use of double search to other multi-objective optimization problems in future work.

\section{Acknowledgement}

The research at the University of California, Irvine, the Carnegie Mellon University and the University 
of Southern California was supported by the National Science Foundation 
(NSF) under grant numbers 2328671, 2441629, 2434916, 2321786, 2112533, and 2121028, as 
well as gifts from Amazon Robotics and the Donald Bren Foundation. The views and conclusions contained in this document
are those of the authors and should not be interpreted as representing
the official policies, either expressed or implied, of the sponsoring
organizations, agencies, or the U.S.  government.

\bibliographystyle{IEEEtran} 
\bibliography{strings,myref}

\begin{thebibliography}{10}
\providecommand{\url}[1]{#1}
\csname url@samestyle\endcsname
\providecommand{\newblock}{\relax}
\providecommand{\bibinfo}[2]{#2}
\providecommand{\BIBentrySTDinterwordspacing}{\spaceskip=0pt\relax}
\providecommand{\BIBentryALTinterwordstretchfactor}{4}
\providecommand{\BIBentryALTinterwordspacing}{\spaceskip=\fontdimen2\font plus
\BIBentryALTinterwordstretchfactor\fontdimen3\font minus
  \fontdimen4\font\relax}
\providecommand{\BIBforeignlanguage}[2]{{%
\expandafter\ifx\csname l@#1\endcsname\relax
\typeout{** WARNING: IEEEtran.bst: No hyphenation pattern has been}%
\typeout{** loaded for the language `#1'. Using the pattern for}%
\typeout{** the default language instead.}%
\else
\language=\csname l@#1\endcsname
\fi
#2}}
\providecommand{\BIBdecl}{\relax}
\BIBdecl

\bibitem{wurman2008coordinating}
P.~R. Wurman, R.~D'Andrea, and M.~Mountz, ``Coordinating hundreds of
  cooperative, autonomous vehicles in warehouses,'' \emph{Artificial
  Intelligence}, vol.~29, no.~1, pp. 9--9, 2008.

\bibitem{stern2019multi}
R.~Stern, N.~Sturtevant, A.~Felner, S.~Koenig, H.~Ma, T.~Walker, J.~Li,
  D.~Atzmon, L.~Cohen, T.~Kumar \emph{et~al.}, ``Multi-agent pathfinding:
  Definitions, variants, and benchmarks,'' in \emph{{Proceedings of the
  International Symposium on Combinatorial Search (SoCS)}}, vol.~10, no.~1,
  2019, pp. 151--158.

\bibitem{wagner2011m}
G.~Wagner and H.~Choset, ``{M*: A complete multirobot path planning algorithm
  with performance bounds},'' in \emph{{Proceedings of the {IEEE/RSJ}
  International Conference on Intelligent Robots and Systems (IROS)}}.\hskip
  1em plus 0.5em minus 0.4em\relax IEEE, 2011, pp. 3260--3267.

\bibitem{sharon2015conflict}
G.~Sharon, R.~Stern, A.~Felner, and N.~R. Sturtevant, ``{Conflict-based search
  for optimal multi-agent pathfinding},'' \emph{Artificial Intelligence}, vol.
  219, pp. 40--66, 2015.

\bibitem{boyarski2015icbs}
E.~Boyarski, A.~Felner, R.~Stern, G.~Sharon, O.~Betzalel, D.~Tolpin, and
  E.~Shimony, ``{ICBS}: The improved conflict-based search algorithm for
  multi-agent pathfinding,'' in \emph{{Proceedings of the International
  Symposium on Combinatorial Search (SoCS)}}, vol.~6, no.~1, 2015, pp.
  223--225.

\bibitem{li2019improved}
J.~Li, A.~Felner, E.~Boyarski, H.~Ma, and S.~Koenig, ``Improved heuristics for
  multi-agent path finding with conflict-based search.'' in \emph{{Proceedings
  of the International Joint Conference on Artificial Intelligence (IJCAI)}},
  vol. 2019, 2019, pp. 442--449.

\bibitem{yu2013structure}
J.~Yu and S.~LaValle, ``{Structure and intractability of optimal multi-robot
  path planning on graphs},'' in \emph{{Proceedings of the AAAI Conference on
  Artificial Intelligence (AAAI)}}, vol.~27, no.~1, 2013, pp. 1443--1449.

\bibitem{erdmann1987multiple}
M.~Erdmann and T.~Lozano-Perez, ``On multiple moving objects,''
  \emph{Algorithmica}, vol.~2, pp. 477--521, 1987.

\bibitem{silver2005cooperative}
D.~Silver, ``{Cooperative pathfinding},'' in \emph{{Proceedings of the AAAI
  Conference on Artificial Intelligence and Interactive Digital Entertainment
  (AIIDE)}}, vol.~1, no.~1, 2005, pp. 117--122.

\bibitem{ma2019searching}
H.~Ma, D.~Harabor, P.~J. Stuckey, J.~Li, and S.~Koenig, ``Searching with
  consistent prioritization for multi-agent path finding,'' in
  \emph{{Proceedings of the AAAI Conference on Artificial Intelligence
  (AAAI)}}, vol.~33, no.~01, 2019, pp. 7643--7650.

\bibitem{chan2023greedy}
S.-H. Chan, R.~Stern, A.~Felner, and S.~Koenig, ``Greedy priority-based search
  for suboptimal multi-agent path finding,'' in \emph{{Proceedings of the
  International Symposium on Combinatorial Search (SoCS)}}, vol.~16, no.~1,
  2023, pp. 11--19.

\bibitem{li2022mapf}
J.~Li, Z.~Chen, D.~Harabor, P.~J. Stuckey, and S.~Koenig, ``Mapf-lns2: fast
  repairing for multi-agent path finding via large neighborhood search,'' in
  \emph{{Proceedings of the AAAI Conference on Artificial Intelligence
  (AAAI)}}, vol.~36, no.~9, 2022, pp. 10\,256--10\,265.

\bibitem{okumura2023lacam}
K.~Okumura, ``Lacam: Search-based algorithm for quick multi-agent
  pathfinding,'' in \emph{{Proceedings of the AAAI Conference on Artificial
  Intelligence (AAAI)}}, vol.~37, no.~10, 2023, pp. 11\,655--11\,662.

\bibitem{barer2014suboptimal}
M.~Barer, G.~Sharon, R.~Stern, and A.~Felner, ``{Suboptimal variants of the
  conflict-based search algorithm for the multi-agent pathfinding problem},''
  in \emph{{Proceedings of the International Symposium on Combinatorial Search
  (SoCS)}}, 2014.

\bibitem{li2021eecbs}
J.~Li, W.~Ruml, and S.~Koenig, ``Eecbs: A bounded-suboptimal search for
  multi-agent path finding,'' in \emph{{Proceedings of the AAAI Conference on
  Artificial Intelligence (AAAI)}}, vol.~35, no.~14, 2021, pp.
  12\,353--12\,362.

\bibitem{pearl1982studies}
J.~Pearl and J.~H. Kim, ``Studies in semi-admissible heuristics,'' \emph{IEEE
  Transactions on Pattern Analysis and Machine Intelligence (PAMI)}, vol.~4,
  pp. 392--399, 1982.

\bibitem{cohen2018anytime}
L.~Cohen, M.~Greco, H.~Ma, C.~Hern{\'a}ndez, A.~Felner, T.~S. Kumar, and
  S.~Koenig, ``Anytime focal search with applications.'' in \emph{{Proceedings
  of the International Joint Conference on Artificial Intelligence (IJCAI)}},
  2018, pp. 1434--1441.

\bibitem{tang2024ita}
Y.~Tang, S.~Koenig, and J.~Li, ``Ita-ecbs: A bounded-suboptimal algorithm for
  combined target-assignment and path-finding problem,'' in \emph{{Proceedings
  of the International Symposium on Combinatorial Search (SoCS)}}, vol.~17,
  2024, pp. 134--142.

\bibitem{tang2023solving}
Y.~Tang, Z.~Ren, J.~Li, and K.~Sycara, ``Solving multi-agent target assignment
  and path finding with a single constraint tree,'' in \emph{International
  Symposium on Multi-Robot and Multi-Agent Systems (MRS)}.\hskip 1em plus 0.5em
  minus 0.4em\relax IEEE, 2023, pp. 8--14.

\bibitem{luna2011push}
R.~J. Luna and K.~E. Bekris, ``Push and swap: Fast cooperative path-finding
  with completeness guarantees,'' in \emph{{Proceedings of the International
  Joint Conference on Artificial Intelligence (IJCAI)}}, 2011, pp. 294--300.

\bibitem{wang2008fast}
K.-H.~C. Wang and A.~Botea, ``Fast and memory-efficient multi-agent
  pathfinding,'' in \emph{{Proceedings of the International Conference on
  Automated Planning and Scheduling (ICAPS)}}, 2008, pp. 380--387.

\bibitem{standley2010finding}
T.~Standley, ``{Finding optimal solutions to cooperative pathfinding
  problems},'' in \emph{{Proceedings of the AAAI Conference on Artificial
  Intelligence (AAAI)}}, vol.~24, no.~1, 2010, pp. 173--178.

\bibitem{standley2011complete}
T.~Standley and R.~Korf, ``Complete algorithms for cooperative pathfinding
  problems,'' in \emph{{Proceedings of the International Joint Conference on
  Artificial Intelligence (IJCAI)}}, 2011, pp. 668--673.

\bibitem{wagner2015subdimensional}
G.~Wagner and H.~Choset, ``{Subdimensional expansion for multirobot path
  planning},'' \emph{Artificial intelligence}, vol. 219, pp. 1--24, 2015.

\bibitem{boyarski2015don}
E.~Boyarski, A.~Felner, G.~Sharon, and R.~Stern, ``Don't split, try to work it
  out: Bypassing conflicts in multi-agent pathfinding,'' in \emph{{Proceedings
  of the International Conference on Automated Planning and Scheduling
  (ICAPS)}}, vol.~25, 2015, pp. 47--51.

\bibitem{li2021pairwise}
J.~Li, D.~Harabor, P.~J. Stuckey, H.~Ma, G.~Gange, and S.~Koenig, ``Pairwise
  symmetry reasoning for multi-agent path finding search,'' \emph{Artificial
  Intelligence}, vol. 301, p. 103574, 2021.

\bibitem{stern2019mapf}
R.~Stern, N.~R. Sturtevant, A.~Felner, S.~Koenig, H.~Ma, T.~T. Walker, J.~Li,
  D.~Atzmon, L.~Cohen, T.~K.~S. Kumar, E.~Boyarski, and R.~Bartak,
  ``{Multi-Agent Pathfinding: Definitions, Variants, and Benchmarks},'' in
  \emph{{Proceedings of the International Symposium on Combinatorial Search
  (SoCS)}}, vol.~10, no.~1, 2019, pp. 151--158.

\end{thebibliography}

\end{document}